\documentclass[conference]{IEEEtran}

\IEEEoverridecommandlockouts
\usepackage{graphicx}
\usepackage{amsmath}
\usepackage{amsfonts}
\usepackage{amssymb}
\usepackage{subfigure}
\usepackage{psfrag}
\usepackage{color}
\usepackage{multicol}
\usepackage{tikz}
\usepackage{verbatimbox}
\usepackage{graphicx}
\usepackage{epsfig}
\usepackage{multirow}
\usepackage{lettrine}
\usepackage{cite}

\usepackage{lipsum,adjustbox}
\usetikzlibrary{calc,positioning,automata}

\newcommand{\bm}{\mathbf}
\newcommand{\bs}{\boldsymbol}
\newcommand{\be}{\begin{equation}}
\newcommand{\ee}{\end{equation}}
\newcommand{\bea}{\begin{eqnarray}}
\newcommand{\eea}{\end{eqnarray}}

\newcommand{\bzero}{{\bm 0}}
\newcommand{\bone}{{\bm 1}}

\newcommand{\ba}{{\bm a}}

\newcommand{\bc}{{\bm c}}

\newcommand{\e}{{\bm e}}

\newcommand{\p}{{\bm p}}

\newcommand{\br}{{\bm r}}
\newcommand{\s}{{\bm s}}
\newcommand{\x}{{\bm x}}
\newcommand{\y}{{\bm y}}
\newcommand{\z}{{\bm z}}

\newcommand{\bvar}{{\bs{\varepsilon}}}

\newcommand{\bd}{{\bf d}}

\newcommand{\bh}{{\bf h}}

\newcommand{\bA}{{\bm A}}
\newcommand{\bB}{{\bm B}}

\newcommand{\bD}{{\bm D}}
\newcommand{\bE}{{\bm E}}
\newcommand{\bF}{{\bm F}}
\newcommand{\bG}{{\bm G}}
\newcommand{\bH}{{\bm H}}
\newcommand{\bI}{{\bm I}}
\newcommand{\bJ}{{\bm J}}
\newcommand{\bP}{{\bm P}}
\newcommand{\bR}{{\bm R}}
\newcommand{\bS}{{\bm S}}

\newcommand{\bX}{{\bm X}}

\newcommand{\bZ}{{\bm Z}}

\newcommand{\I}{{\bm I}}

\newcommand{\BD}{{\boldsymbol{\mathcal D}}}

\newcommand{\bLambda}{\mbox{\boldmath$\Lambda$}}

\newcommand{\bGamma}{\mbox{\boldmath$\Gamma$}}
\newcommand{\bOmega}{\mbox{\boldmath$\Omega$}}
\newcommand{\bPsi}{\mbox{\boldmath$\Psi$}}

\newcommand{\bPhi}{\mbox{\boldmath{$\Phi$}}}

\newcommand{\bPi}{\mbox{\boldmath{$\Pi$}}}

\title{Synchronization for Multiuser Uplink OTFS \vspace{-0.3cm}}
\author{\IEEEauthorblockN{Mohsen Bayat, Sanoopkumar P.S., and Arman Farhang} 
\IEEEauthorblockA{Department of Electronic \& Electrical Engineering, Trinity College Dublin, Ireland \\ \{bayatm, pungayis, arman.farhang\}@tcd.ie}
\vspace{-1cm}
\thanks{
This publication has emanated from research conducted with the financial support of Taighde Éireann – Research Ireland under Grant number 19/FFP/7005(T) and 21/US/3757.
}
}

\begin{document}
\maketitle
\begin{abstract}
In this paper, we propose time and frequency synchronization techniques for the uplink of multiuser orthogonal time frequency space (MU-OTFS) in high-mobility scenarios. We introduce a spectrally efficient and practical pilot pattern where each user utilizes a pilot with a cyclic prefix (PCP) within a shared pilot region on the delay-Doppler plane. At the receiver, a bank of filters is deployed to separate the users' signals and accurately estimate their timing offsets (TOs) and carrier frequency offsets (CFOs). Our technique employs a threshold-based approach that provides precise TO estimates. Our proposed CFO estimation technique reduces the multi-dimensional maximum likelihood (ML) search problem into multiple one-dimensional search problems. Furthermore, we apply the Chebyshev polynomials of the first kind basis expansion model (CPF-BEM) to effectively handle the time-variations of the channel in obtaining the CFO estimates for all the users. Finally, we numerically investigate the error performance of our proposed synchronization technique in high mobility scenarios for the MU-OTFS uplink. Our simulation results confirm the efficacy of the proposed technique in estimating the TOs and CFOs which also leads to an improved channel estimation performance.   
\end{abstract}

\vspace{-0.1cm}
\section{introduction}\label{sec:Introduction}
The sixth-generation networks (6G) is expected to support a variety of applications in high-mobility and high-frequency environments. In such scenarios, wireless channels become doubly selective and orthogonal frequency division multiplexing (OFDM) as a technology of choice in the existing wireless standards loses its benefits \cite{Yang2019}. 
To overcome this challenge, orthogonal time frequency space (OTFS) modulation was introduced in \cite{Hadani2017}. OTFS operates in the delay-Doppler domain where the wireless channel has a sparse and time-invariant representation \cite{Wei2021}.
Since its emergence, there has been growing interest among researchers in this technology, as highlighted by the formation of a large body of literature in this area \cite{Raviteja2018}.
However, OTFS is still in its early stages of development as the majority of its literature is focused on single-user systems. 
This is while multiple access and multiuser capabilities are important aspects of the future wireless networks \cite{Wei2021}. 

Various orthogonal multiple access techniques for OTFS are discussed in \cite{Augustine2019, Khan2019, Farhang2024, Raviteja2018}. Specifically, interleaved resource allocation in the delay-Doppler and time-frequency domains were proposed in \cite{Augustine2019} and \cite{Khan2019}, respectively. However, these schemes have limitations. To overcome these limitations, a generalized resource allocation scheme was introduced in \cite{Farhang2024}.
For channel estimation of multiuser OTFS (MU-OTFS), a pilot pattern was proposed in \cite{Raviteja2018} where the users' pilots are multiplexed along the delay dimension. However, the required delay guards between the users' pilots leads to spectral efficiency loss.
While the works in \cite{Augustine2019, Khan2019, Farhang2024, Raviteja2018} address different aspects of MU-OTFS, as of yet, to the best of our knowledge, there is no comprehensive literature on synchronization in multiuser uplink scenario. 
Synchronization errors can adversely affect the subsequent blocks of the receivers resulting in performance degradation. Hence, the focus of this paper is on synchronisation of OTFS in the uplink.
Multiple timing offsets (TOs) and carrier frequency offsets (CFOs) in MU-OTFS lead to interference between the users' pilot signals. Due to such interference, the single-user synchronization techniques are not adaptable to multiuser scenarios.
Except \cite{Sinha2020}, the existing literature on OTFS synchronization is focused on single-user scenarios \cite{Das2021, Saquib2021, Saquib2024, Chung2024, Xu2023, Sun2024, SLi2024, Bayat2022, Bayat2023}.
In \cite{Sinha2020}, the authors propose a closed-loop TO estimation technique, where the estimated TO is fed back to the mobile terminal (MT). However, such an approach may lead to outdated TO estimates in the next uplink transmission. Thus, an open-loop solution can effectively address this issue.
In \cite{Das2021}, TO and CFO are estimated as a part of the channel in single-user scenario. 
However, this approach is not applicable to the uplink as accurate estimation of multiple TOs is required to locate the pilots of the users.
Moreover, as we show in this paper, absorbing the CFOs into the channel leads to channel estimation performance degradation. To the best of our knowledge, these problems have not been reported in the literature to date.

To address the above challenges, in this paper, we propose a novel pilot pattern for MU-OTFS uplink while we develop effective TO and CFO estimation techniques. In our proposed pilot pattern, all the users share a common pilot region on the delay-Doppler plane to avoid spectral efficiency loss. To better understand the effects of multiuser synchronization errors on OTFS, we formulate the received MU-OTFS signal in the uplink. Then, we analyze the effect of CFOs on the Doppler spectrum. We derive an input-output relationship for the received uplink signal with a compound channel matrix that encompasses the TOs, CFOs, and the channels of all the users. As the uplink signal is affected by the users' TOs and CFOs, aligning one user's signal in time and frequency leads to misalignment of the other users' signals. To tackle this issue, we show that our derived compound channel representation can be used to jointly compensate the TOs and CFOs of all the users, after they are estimated.
For effective estimation of the users' TOs and CFOs, we propose deploying a bank of filters to separate their signals and alleviate the effect of interference between the users' pilots.
We propose a threshold-based method that provides highly accurate TO estimates.
Then, we develop a maximum likelihood (ML) estimation technique for frequency synchronization. 
User separation by the bank of filters reduces the multidimensional search for estimating multiple CFOs into multiple single-dimensional search procedures in parallel.
In this technique, we employ the Chebyshev polynomial-exponential basis expansion model (CPF-BEM) to absorb the time variation of channel coefficients into the basis function and effectively estimate the CFOs.
To analyze the performance of our proposed techniques, we numerically examine the mean and variance of the TO estimation error, and the mean squared error (MSE) of the CFO estimates versus signal-to-noise ratio (SNR) and normalized Doppler spread.
Our simulations confirm the efficacy of our proposed techniques in providing accurate TOs and CFOs which also leads to an improved channel estimation performance.

\textit{Notations}:
Scalars, vectors, and matrices are denoted by regular, bold lowercase, and bold uppercase letters, respectively. $\mathbb{C}^{M \times N}$ represents the set of $M \times N$ complex matrices, while $\bm{I}_{N}$, $\bm{1}_{N}$, and $\bm{0}_{N}$ refer to $N \times N$ identity, one, and zero matrices.
The Hermitian, transpose, and inverse are represented as, $(.)^{\rm{H}}$, $(.)^{\rm{T}}$, and $(.)^{-1}$, respectively.
The element-wise product, Kronecker product, and Dirac delta function are denoted by $\odot$, $\otimes$, and $\delta[\cdot]$, respectively.
$\bm{F}_N$ is the normalized $N$-point discrete Fourier transform (DFT) matrix with the elements $F[l,n] = \frac{1}{\sqrt{N}} e^{-j \frac{2 \pi ln}{N}}$ for $l,n=0,\ldots,N-1$.
Operators $(\!(.)\!)_i$, $(.)_i$, ${\rm{vec}} \{ . \}$, ${\rm{diag}}[.]$, and $\max_i \{ . \}$ represent a circular shift, modulo, vectorization, diagonal matrix, and the maximum of a function with respect to $i$, respectively.
Finally, $\bm{C} = {\rm{circ}} \{ \bm{c} \}$ represents a circulant matrix whose first column is $\bm{c}$.
\vspace{-0.2cm}
\section{System Model} \label{sec:System}
\vspace{-0.1cm}
In this paper, we consider an uplink OTFS system where $Q$ users are simultaneously communicating with the base station (BS) simultaneously. At each MT, the information bits are mapped onto quadrature amplitude modulation (QAM) constellation and placed on a delay-Doppler grid with $M$ delay bins and $N$ Doppler bins.
We present the sets of $M_q$ delay bins and $N_q$ Doppler bins allocated to the  
users $q=0,\ldots,Q-1$ as $\mathbb{U}_\tau^q$ and $\mathbb{U}_\nu^q$, respectively. The delay-Doppler resources allocated to any pair of distinct users $q,p\in\{1,\ldots,Q \}$ belong to mutually exclusive sets $U_{\tau/\nu}^q \cap U_{\tau/\nu}^p=\emptyset$ where $\cup_{q=1}^Q \mathbb{U}_\tau^q=\{0,\ldots,M-1\}$ and $\cup_{q=1}^Q \mathbb{U}_\nu^q=\{0,\ldots,N-1\}$.
After stacking the QAM symbols of user $q$ into $M_q\times N_q$ data matrix $\BD^q$, they are mapped onto their corresponding delay-Doppler bins. This is done by the delay and Doppler resource allocation matrices $\bGamma_\tau^q$ and $\bGamma_\nu^q$, respectively,
to obtain the $M \times N$ matrix $\bD^q$ as $\bD^q = \bGamma_{\tau}^q \BD^q \bGamma_{\nu}^q$ with the elements $D^q[l,k]$ for $l=0,\ldots, M-1$ and $k=0,\ldots, N-1$.
The matrices $\bGamma_\tau^q$ and $\bGamma_\nu^q$ are formed by the columns of $\I_{M}$ with the indices in $\mathbb{U}_\tau^q$ and the rows of  $\I_{N}$ with the indices from the set $\mathbb{U}_\nu^q$, respectively.
Similar to \cite{Farhang2024}, we consider generalized resource allocation.

Then, the transmit signal of user $q$ is formed by converting the data symbols into the delay-time domain via an inverse DFT (IDFT) operation along the Doppler dimension, i.e., as
\begin{align} \label{eqn:ifft} \bX^q = \bD^q  \bF_N^{\rm{H}}, \end{align}
with the elements of $\bX^q$ represented as
$X^q[l,n] = \frac{1}{\sqrt{N}} \sum_{k=0}^{N-1} D^q[l,k] e^{j \frac{2 \pi kn}{N}}$ for time index $n=0,\ldots,N-1$.
After parallel-to-serial conversion, $\x^q = {\rm{vec}} \{ \bX^q \}$, a CP of length $L_{\rm{cp}}$ is appended at the beginning of the OTFS frame for user $q$ as
\begin{align} \label{eqn:Acp} \s^q = \bA_{\rm{cp}}  \x^q, \end{align}
where $\bA_{\rm{cp}}=[\bJ_{\rm{cp}}^{\rm{T}},\bI^{\rm{T}}_{MN}]^{\rm{T}}$ is the CP addition matrix and the matrix $\bJ_{\rm{cp}}$ includes the last $L_{\rm{cp}}$ rows of the identity matrix $\bI_{MN}$. We consider a quasi-synchronous system in time where the signals from MTs arriving at the BS are time-aligned within the CP \cite{Morelli2007}.
In a quasi-synchronous system, the channel length for user $q$ extends from $L^q_{\rm{ch}}$ to $\check{L}^q_{\rm{ch}}\!=\!L^q_{\rm{ch}}+\theta^q$, where $\theta^q$ is the TO of user $q$, normalized by the delay spacing.
Thus, the CP length can be chosen as $L_{\rm{cp}} \!\!=\!\! \max_q \{ \check{L}^q_{\rm{ch}} \} \!=\! \max_q \{ L^q_{\rm{ch}} \} \!+\! \theta_{\rm{max}} \!-\! 1$, where $\theta_{\rm{max}}$ is the maximum delay spread normalized to the delay spacing. 
Substituting (\ref{eqn:ifft}) into (\ref{eqn:Acp}), the transmit signal can be obtained as $\s^q \!=\! \bA_{\rm{cp}} (\bF_N^{\rm{H}} \!\otimes\! \bI_M) \bd^q$
where $\bd^q \!\!=\! \bGamma^q \check{\bd}^q$, $\check{\bd}^q \!\!=\! {\rm vec}\{\BD^q\}$ and $\bGamma^q \!=\! (\bGamma^{q}_{\nu})^{\rm{T}} \otimes \bGamma_\tau^q$.

Considering multiple TO and CFO effects of the received signal from all MTs at the BS can be represented as
\begin{align} \label{eqn:rec20}
r[\kappa] &= \sum_{q=0}^{Q-1} e^{j \frac{2 \pi \varepsilon^q}{N_{\rm{s}}}\kappa} \sum_{\ell=0}^{L^q_{\rm{ch}}-1} s^q[\kappa-\ell-\theta^q] h^q[\ell,\kappa] + \eta[\kappa],
\end{align}
where $\kappa=0,\ldots,N_{\rm{s}}-1$ and $N_{\rm{s}}=MN+L_{\rm{cp}}$ is the total number of samples for an OTFS frame over a duration of $T$ seconds.
The parameter $\varepsilon^q$ represents the CFO of user $q$, normalized by the Doppler spacing.
In (\ref{eqn:rec20}), $h^q[\ell,\kappa]$ is the impulse response of the linear time-varying (LTV) channel between the user $q$ and BS antenna, expressed as
\be \label{eqn:ch0} h^q[\ell,\kappa]=\sum_{i=0}^{\Upsilon^q-1} h^q_i e^{j 2 \pi \nu_i^q (\kappa-\ell)} \delta[\ell-\ell^q_i], \ee
where $h^q_i$, $\ell^q_i$, and $\nu^q_i=\nu^q_{\rm{max}} \cos \psi^q_i$, are channel gains, delays, maximum Doppler shifts, and the angle of arrival, respectively, for path $i$ and user $q$.
Additionally, $\Upsilon^q$ is the total number of paths and $\eta[\kappa] \!\sim\! \mathcal{CN} (0,\sigma^2_\eta)$ is the complex additive white Gaussian noise (AWGN) with the variance $\sigma^2_\eta$.
By substituting (\ref{eqn:ch0}) into (\ref{eqn:rec20}), the received signal can be rearranged as
\begin{align} \label{eqn:rec2}
r[\kappa]
&= \sum_{q=0}^{Q-1} \sum_{\ell=0}^{L^q_{\rm{ch}}-1} \check{h}^q[\ell,\kappa] s^q[\kappa-\ell-\theta^q] + \eta[\kappa],
\end{align}
where $\check{h}^q[\ell,\kappa]=\sum_{i=0}^{\Upsilon^q-1} \check{h}^q_i e^{j 2 \pi \check{\nu}_i^q (\kappa-\ell)} \delta[\ell-\ell^q_i]$ with $\check{h}^q_i=h^q_i e^{j 2 \pi \delta^q \ell}$, $\check{\nu}^q_i=\delta^q+\nu_i^q$, and $\delta^q=\frac{\varepsilon^q}{N_{\rm{s}}}$.

Equation (\ref{eqn:rec2}) shows that the CFO of each user is added as a constant to all the Doppler shifts of its channel, increasing the maximum Doppler frequency of the channel, \cite{Das2021}. 
This effect makes channel estimation more challenging, especially as $|\varepsilon^q|$ increases.
From (\ref{eqn:rec2}), one may realise that the CFO and TO for each user can be absorbed into its channel and thus, can be estimated as a part of the channel and compensated at the equalization stage.
The TOs though need to be compensated so that we can find the location of the pilot for each user on the delay-Doppler plane.
As shown in Section~\ref{sec:Result}, if the CFOs are estimated as a part of the channel, the channel estimates are not as accurate as estimating and compensating the CFOs first and then estimating the channel.
Hence, in the following sections, we propose a CFO estimation technique for the MU-OTFS uplink.
Before proposing the techniques for estimating TO and CFO values, let us first investigate the impact of these offsets on the equivalent channel in the delay-Doppler domain. To understand the real effects of TO and CFO on the MU-OTFS channel, we derive the compound channel model for an asynchronous MU-OTFS system in the following section.

\vspace{-0.1cm}
\section{Channel Effect}\label{sec:Ch}
\begin{figure}[!t]
  \centering 
  {\includegraphics[scale=0.185]{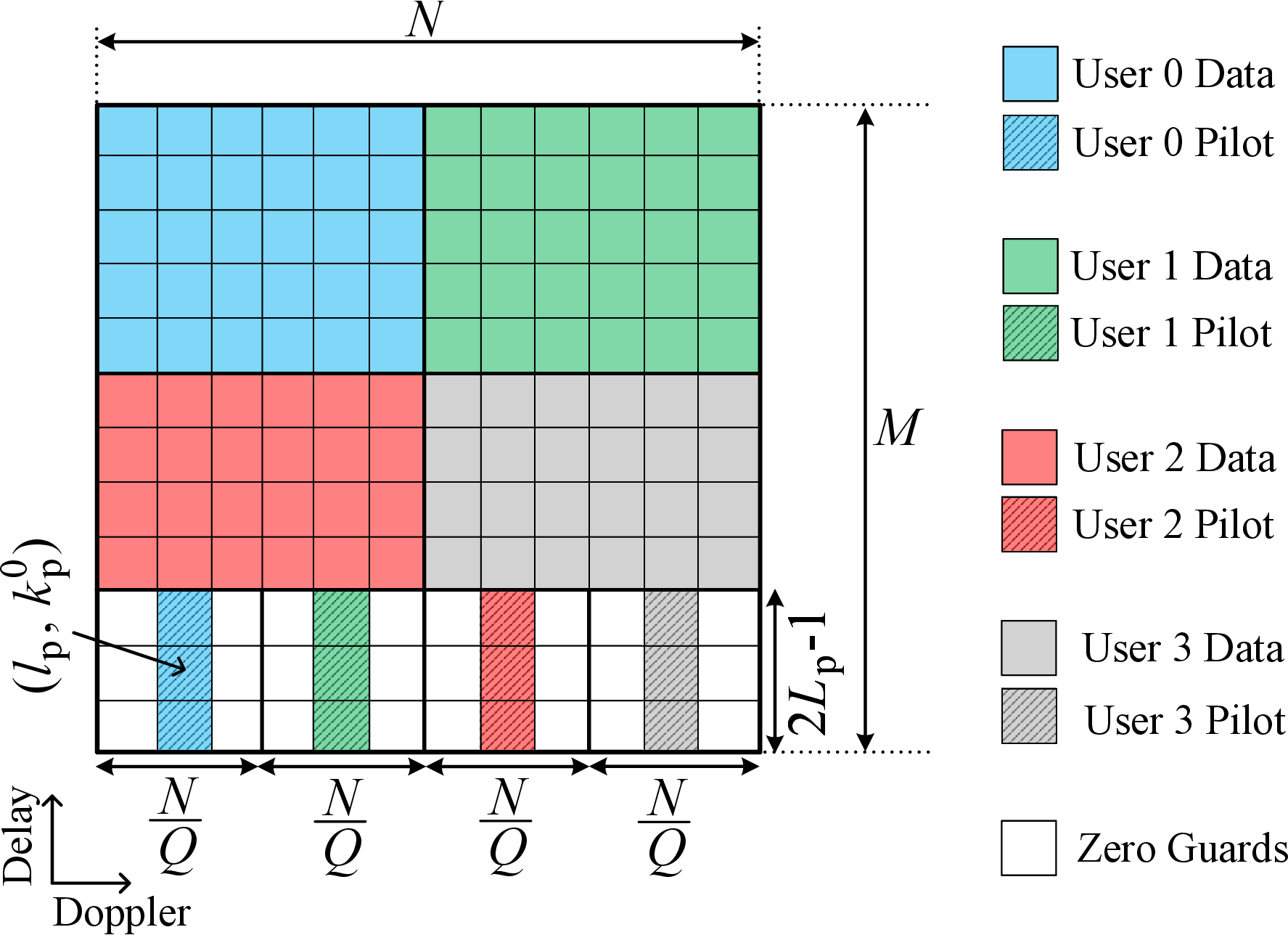}}
  \vspace{-0.3cm}
  \caption{Proposed PCP structure for MU-OTFS in the delay-Doppler domain.}
  \vspace{-0.4cm}
  \label{fig:pilot}
\end{figure}
Let us consider the received signal at the BS from all MTs after transmission over the LTV channel by stacking the values of $r[\kappa]$, $s^q[\kappa]$ and $\eta[\kappa]$ for $\kappa=\theta_{\rm{max}},\ldots,MN+L_{\rm{cp}}-1$ into the vectors $\br$, $\s^q$ and $\boldsymbol{\eta}$, and using (\ref{eqn:rec20}) and (\ref{eqn:rec2}), the received signal vector can be represented as
\begin{align} \label{eqn:rec} \br &= \sum_{q=0}^{Q-1} \bPhi(\varepsilon^q) \bPi(\theta^q) \bH^q \s^q + \boldsymbol{\eta}
, \end{align}
where
$\bPhi(\varepsilon^q)={\rm{diag}} [ e^{j \frac{2\pi \varepsilon^q (\theta_{\rm{max}})}{N_{\rm{s}}}},\ldots,e^{j \frac{2\pi \varepsilon^q (MN+L_{\rm{cp}}-1)}{N_{\rm{s}}}} ]$ and
$\bPi(\theta^q)=[\mathbf{0}_{(MN+\check{L}_{\rm{cp}}) \times \theta^q},\bI_{(MN+\check{L}_{\rm{cp}})},\mathbf{0}_{(MN+\check{L}_{\rm{cp}}) \times (\theta_{\rm{max}}-\theta^q)}]$
are the CFO and TO matrices, respectively. 
The $\bH^q \in \mathbb{C}^{N_{\rm{s}}}$ is a Toeplitz matrix that stacks the channel coefficients from (\ref{eqn:ch0}).
After removing the CP of length $\check{L}_{\rm{cp}}=L_{\rm{cp}}-\theta_{\rm{max}}$, the received signal can be expressed as $\y=\bR_{\rm{cp}} \br$ where  the CP removal matrix is defined as $\bR_{\rm{cp}}=[\mathbf{0}_{MN\times \check{L}_{\rm{cp}}},\bI_{MN}]$.
Hence, the received delay-Doppler domain signal is obtained by applying $N$-point DFT to $\y$ along the time dimension as 
\begin{align} \label{eqn:d_hat} 
\widetilde{\bd} \!&=\! (\bF_N \!\otimes\! \bI_M) \y \!=\!\!\! \sum_{q=0}^{Q-1} (\bF_N \otimes \bI_M) \check{\bPhi}(\varepsilon^q) {\bLambda}^q (\bF_N^{\rm{H}} \otimes \bI_M) \bd^q \!+\! \check{\boldsymbol{\eta}} \nonumber \\ &= \sum_{q=0}^{Q-1} \check{\bPhi}_{\rm{DD}}(\varepsilon^q) \bLambda^q_{\rm{DD}} \bd^q + \check{\boldsymbol{\eta}} = \bPsi_{\rm{DD}} \bd + \check{\boldsymbol{\eta}}.
\end{align}
In the first line of (\ref{eqn:d_hat}), the CFO matrix of each user after CP removal is presented by $\check{\bPhi}(\varepsilon^q)\!\!=\!\!{\rm{diag}}[e^{j \frac{2\pi \varepsilon^q (L_{\rm{cp}})}{N_{\rm{s}}}}\!\!\!,\!\ldots\!,\!e^{j \frac{2\pi \varepsilon^q (MN+L_{\rm{cp}}-1)}{N_{\rm{s}}}}]$, ${\bLambda}^q \!=\! \bR_{\rm{cp}} \bPi(\theta^q) \bH^q \bA_{\rm{cp}}$ is the combination of TO and channel response for user $q$, and $\check{\boldsymbol{\eta}}=(\bF_N \otimes \bI_M) \bR_{\rm{cp}} \boldsymbol{\eta}$ is AWGN in the delay-Doppler domain.
In (\ref{eqn:d_hat}), $\check{\bPhi}_{\rm{DD}}(\varepsilon^q) = (\bF_N \otimes \bI_M) \check{\bPhi}(\varepsilon^q) (\bF^{\rm{H}}_N \otimes \bI_M)$ and $\bLambda^q_{\rm{DD}}=(\bF_N \otimes \bI_M) {\bLambda}^q (\bF_N^{\rm{H}} \otimes \bI_M)$ are the CFO matrix and the channel response of the $q^{\text{th}}$ user in the delay-Doppler domain, respectively.
Ultimately, $\bPsi_{\rm{DD}}\!=\!\sum_{q=0}^{Q-1} \! \check{\bPhi}_{\rm{DD}}(\varepsilon^q) \bLambda^q_{\rm{DD}} (\bGamma^q)^{\rm{H}}$ is the compound channel that includes the TO, CFO, and channel responses of all the users and $\bd=\sum_{q=0}^{Q-1} \bd^q = \sum_{q=0}^{Q-1} \bGamma^q \check{\bd}^q$ is the combined delay-Doppler domain transmitted data symbols of all the users.

From (\ref{eqn:d_hat}), one might deduce that the best way to compensate the effects of multiple TOs and CFOs in the uplink and equalize the channel is by using the compound channel matrix $\bPsi_{\rm{DD}}$.
However, as previously mentioned,
channel estimation after synchronization is more accurate than directly estimating the channel which includes the CFO and TO effects.
This is because TO estimation is required to find the users' pilot signals and CFO compensation before channel estimation prevents the unwanted channel estimation errors.
\vspace{-0.05cm}
\section{Proposed Synchronization Technique} \label{sec:ML}
In a multiuser uplink system, the estimated TOs for each user are fed back to the user for pre-compensation in the next uplink transmission \cite{Morelli2007}. However, in an OTFS system, the TO estimates are required for locating the pilot region which is used for the CFO and channel estimation \cite{Bayat2022}.
In this section, we address the problem of multiple TO estimation and then, we jointly estimate the CFOs and channels of all the users.
We propose a designated pilot region where all the users transmit their pilots and the remaining delay-Doppler bins are arbitrarily allocated to the users for data transmission, see Fig.~\ref{fig:pilot}.
To estimate these parameters, we deploy the pilot with a cyclic prefix (PCP) as a pilot for synchronization that was originally proposed for channel estimation in \cite{Sanoop2023}.
We use PCP as it is a practical pilot opposed to the impulse pilot that has prohibitively large peak-to-average power ratio (PAPR).
For each user, a PCP of length $2L_{\rm{p}}-1$ is formed a by Zadoff-Chu (ZC) sequence of length $L_{\rm{p}}$ in which the last $L_{\rm{p}}-1$ elements of the sequence are copied at the beginning as a CP.
In our proposed pilot structure, the pilots of the users $q=0,\ldots,Q-1$ reside within the same delay region, i.e.,  $l = l_{\rm p}-L_{\rm p}+1,\ldots, l_{\rm p}+L_{\rm p}-1$ in Doppler bins $ k^q_{\rm{p}}=k_{\rm{p}} + q \lfloor \frac{N}{Q} \rfloor$ for $k_{\rm{p}}=0,\ldots,\lfloor \frac{N}{Q} \rfloor-1$, see Fig.~\ref{fig:pilot}.

\begin{figure}[!t]
  \centering 
  {\includegraphics[scale=0.265]{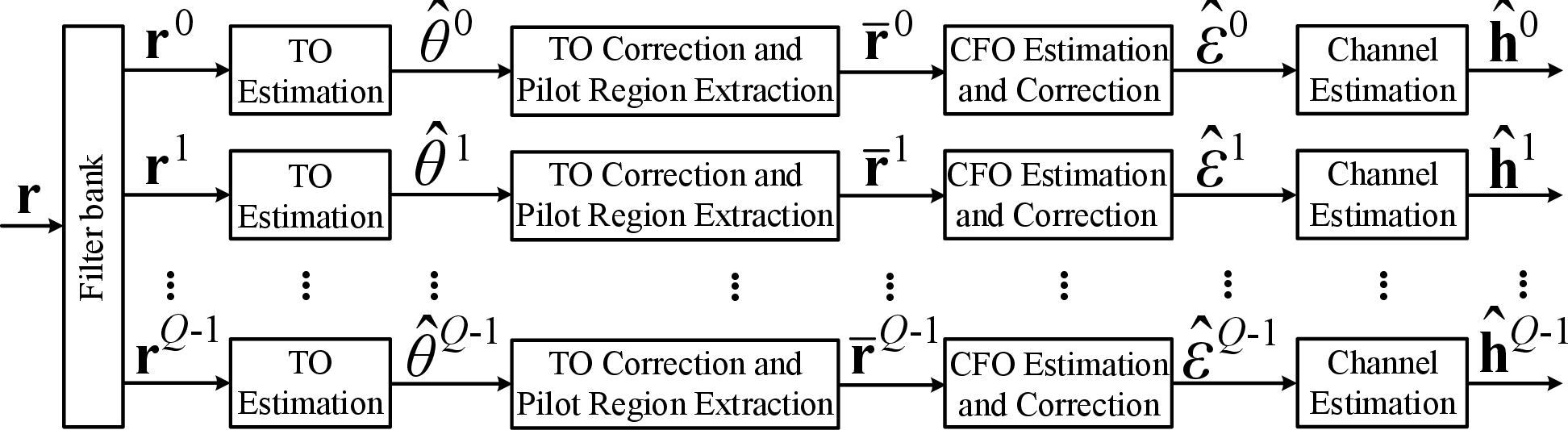}}
  \vspace{-0.1cm}
  \caption{The proposed structure for TO, CFO, and Channel estimation using a bank of bandpass filters.}
  \vspace{-0.4cm}
  \label{fig:FB}
\end{figure}

\subsection{Proposed Correlation-Based TO Estimation Technique} \label{sec:TO}
As mentioned earlier, multiple TO estimations in the uplink of OTFS are necessary for locating users' pilot signals to perform CFO and channel estimation.
Furthermore, TO estimation in the uplink enables the BS to feed the estimated TOs back to the MTs for pre-compensation before the next transmission period.
As we do not estimate the TOs as a part of the channel, the parameter $L_{\rm{p}}$ only depends on $\max_q\{L_{\rm{ch}}^q\}$. This is an important aspect of our approach as it requires a shorter pilot than the case where TO would be absorbed into the channel. This makes our pilot structure spectrally efficient.

To estimate the TO of each user, we propose to pass the received signal $\br$ through a bank of digital bandpass filters that separates different users' pilots, see Fig.~\ref{fig:FB}.
Considering the brickwall filter $\ba^q = (\!(\ba)\!)_{q \lfloor \frac{N}{Q} \rfloor }$ where $\ba=[\bone^{\rm{T}}_{\lfloor \frac{N}{Q} \rfloor \times 1},\bzero^{\rm{T}}_{(Q-1)\lfloor \frac{N}{Q} \rfloor \times 1}]^{\rm{T}}$, the signal of a given user $q$ can be separated by multiplication of the received signal in Doppler by the filter response which is equivalent to circular convolution in delay-time domain by $\e^q = \bF^{\rm{H}}_N \ba^q$. Hence, circular convolution matrix $\bE^q={\rm{circ}} \{ \e^q \}$ can be used as follows to obtain $\br^q$.
This enables the BS to independently estimate the TOs for different users and thus, reduce a $Q$-dimensional search problem into $Q$ parallel one-dimensional search problems. 
The separated pilot signal for a given user $q$ can be expressed as
\be \label{eqn:BF} \br^q = \big( (\bE^q)^{\rm{H}} \otimes \bI_M \big) \check{\br}, \ee
where $q=0,\ldots,Q-1$ and $\check{\br}$ represents the first $MN$ elements of the received vector $\br$.

\begin{figure}[!t]
  \centering 
  {\includegraphics[scale=0.172]{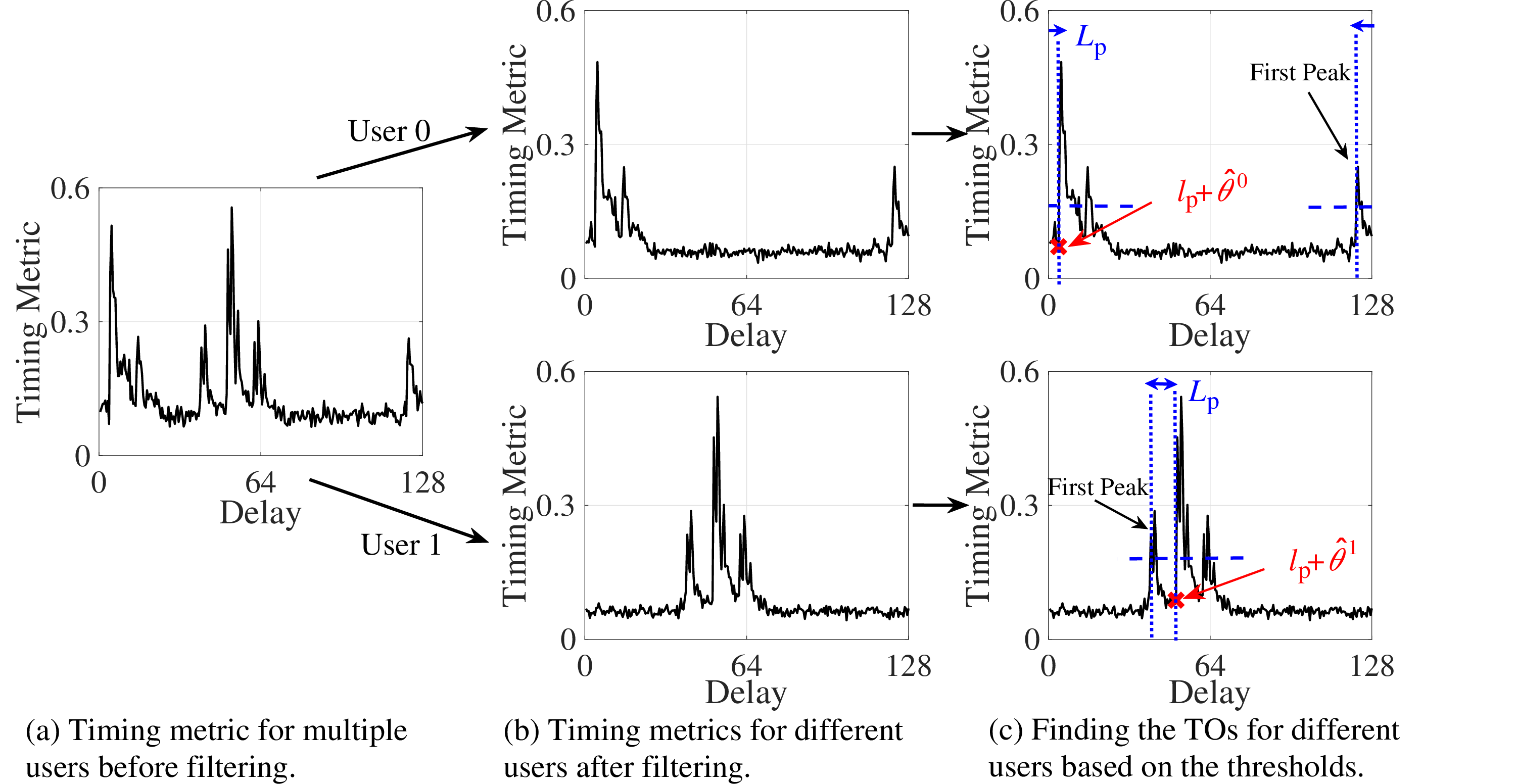}}
  \vspace{-0.3cm}
  \caption{A snapshot of the timing metrics for $Q=2$ and $\nu^q_{\rm{max}} T \approx 2.91 \, \forall q$.}
  \vspace{-0.3cm}
  \label{fig:Mean}
\end{figure}

After the separation of different pilot signals of the users, the TOs for different users can be estimated in parallel by finding the two identical halves of the pilot in the delay-time domain. 
Hence, the transmitted pilot of each user in the delay-time dimension can be correlated and slid over the signal $\br^q$ to estimate $\theta^q$ for $q=0,\ldots,Q-1$, see Fig.~\ref{fig:Mean}-(a) and -(b).
The received vector $\br^q$ can be used to form $\bR^q [l,n] \in \mathbb{C}^{M \times N}$ for $l=0,\ldots,M-1$ and $n=0,\ldots,N-1$.
Additionally, matrix $\bZ^q_{l'}[l,n] \in \mathbb{C}^{M \times N}$ represents the transmitted delay-time pilot signal
$\bZ^q=[\bzero_{(l_{\rm{p}}-L_{\rm{p}}+1) \times N},(\overline{\bZ}^q \bF_N^{\rm{H}}),\bzero_{(M-l_{\rm{p}}-L_{\rm{p}}) \times N}]^{\rm{T}}$,
circularly shifted upwards by $l'$ positions
along the delay dimension.
Given $\overline{\bZ}^q=[\bzero_{(2L_{\rm{p}}-1) \times (k^q_{\rm{p}}\!-1)},\z_{\rm{p}},\bzero_{ (2L_{\rm{p}}-1) \times (N-k^q_{\rm{p}})}]$ and 
$\z_{\rm{p}}$ is a vector of size $(2L_{\rm{p}}-1) \times 1$, containing the PCP sequence elements.
The two-dimensional correlation function for user $q$ can be obtained as
\be \label{eqn:TO2_Corr} \bP^q [l',n] = \frac{1}{M} \Big| \sum_{l=0}^{M-1} \bR^q[l,n] \odot (\bZ^q_{l'}[l,n])^{\rm{H}} \Big|, \ee
where $l'=0,\ldots,M-1$.
Then, the timing metric for TO estimation can be obtained as
\be \label{eqn:tm}
\p^q [l'] = \frac{1}{N} \sum_{n=0}^{N-1} \bP^q[l',n].
\ee

As explained in \cite{Bayat2022} and \cite{Bayat2023}, the peak of the correlation function for TO estimation is primarily influenced by the strongest tap of the channel. Consequently, estimation errors occur when the first tap of the channel is not the dominant one.
Hence, to address this issue, we propose to take a similar approach to \cite{Farhang2024} and find the first peak of $\p^q[l']$ which indicates the first tap of the channel. However, unlike the impulse pilot-based method in  \cite{Farhang2024}, interference between the samples of PCP hinders the effective detection of the first peak of the correlation function. 
Therefore, in this paper, we use the cross-correlation function of the received signal and the pilot signal as described in (\ref{eqn:TO2_Corr}).
To find the first peak, we group all the peaks 
using a threshold value $\mathcal{T}^q$, such that $0<\mathcal{T}^q\leq1$. The constructed set of peaks can be described mathematically as
\be \label{eqn:TO_d} 
\boldsymbol{\widehat{\Theta}}^q \!=\! \Big\{ l' \, \Big| \, \big|\p^q[l'] \big| \!\geq\! \big( \mathcal{T}^q \times \max \big\{ |\p^q[l']| \big\} \big) \Big\}.
\ee
Consequently, the first peak is identified as
\be \label{eqn:TO_d2} 
\hat{\theta}^q= \big( \min \{ \boldsymbol{\widehat{\Theta}}^q \}+L_{\rm{cp}}-l_{\rm{p}}-1 \big)_M.
\ee

After estimating the TOs for different users, the pilot region for each user is identified. In the next stage, 
the CFOs of different users are estimated by $Q$ parallel ML search procedures as it will be explained in the next subsection.

\vspace{-0.15cm}
\subsection{Proposed ML-Based CFO Estimation Technique} \label{sec:CFO}
\vspace{-0.05cm}
As the first step for CFO estimation, we stack the received pilot signal of user $q$ into a vector $\overline{\br}^q$.
Hence, the received samples after removing the CP from the pilot in the delay bins $\check{l}^q_{\rm{p}} = l_{\rm{p}} + \theta^q$ to $\check{l}^q_{\rm{p}} + L_{\rm{p}}-1$ over all the time slots are stacked in the vector $\overline{\br}^q\!=\![(\overline{\br}^q_0)^{\rm T}\!,\ldots\!,\!(\overline{\br}^q_{N\!-\!1})^{\rm T}]^{\rm{T}} \in \mathbb{C}^{NL_{\rm{p}}\times 1}$
where $\overline{\br}^q_n=[r^q[L_{\rm{cp}}+nM+\check{l}^q_{\rm{p}}],\ldots,r^q[L_{\rm{cp}}+nM+\check{l}^q_{\rm{p}}+L_{\rm{p}}-1]]^{\rm T}$.
Using (\ref{eqn:rec}) and (\ref{eqn:BF}), $\overline{\br}^q$ can be expressed as, 
\be \label{eqn:Mtx_rec1} 
\overline{\br}^q = \overline{\bPhi}(\varepsilon^q) \bOmega^q \overline{\s}^q + \overline{\boldsymbol{\eta}}^q, 
\ee
where 
$\overline{\s}^q$ represents the transmitted delay-time pilot for user $q$ as $\overline{\s}^q=[(\overline{\s}^q_0)^{\rm T},\ldots,(\overline{\s}^q_{N-1})^{\rm T}]^{\rm{T}} \in \mathbb{C}^{N L_{\rm{p}} \times 1}$, and $\overline{\s}^q_n$ is the pilot samples of the user $q$ in the delay bins $l_{\rm{p}}$ to $l_{\rm{p}}+L_{\rm{p}}-1$ and the time slot $n$.
Additionally, the CFO matrix in the pilot region is given by $\overline{\bPhi}(\varepsilon^q)\!=\!{\rm{diag}} [\overline{\bPhi}_0(\varepsilon^q),\ldots,\overline{\bPhi}_{N-1}(\varepsilon^q)] \in \mathbb{C}^{N{L}_{\rm{p}}}$ and $\overline{\bPhi}_n(\varepsilon^q)\!\!=\!
{\rm{diag}} \big[ e^{j \frac{2\pi \varepsilon^q (\check{L}_{\rm{cp}}\!+nM+\check{l}^q_{\rm{p}})}{N_{\rm{s}}}}\!\!\!,\ldots\!,e^{j \frac{2\pi \varepsilon^q (\check{L}_{\rm{cp}}\!+nM+\check{l}^q_{\rm{p}}\!+L_{\rm{p}}\!-\!1)}{N_{\rm{s}}}} \big]$.
The channel matrix for user $q$ in the pilot region that realizes the convolution operation can be represented by
${\bOmega}^q = {\rm{diag}}[\bOmega^q_0,\ldots,\bOmega^q_{N-1}] \in \mathbb{C}^{N L_{\rm{p}}}$ where $\bOmega^q_n$ has a structure that is shown in (\ref{eqn:ch2}), on the top of the next page.
Finally, $\overline{\boldsymbol{\eta}}^q \in \mathbb{C}^{NL_{\rm{p}} \times 1}$ is the noise vector in the pilot region of user $q$.
By interchanging the order of convolution in (\ref{eqn:Mtx_rec1}), the received pilot in the time slot $n$, can be expressed as
\begin{align}
\setcounter{equation}{14} \label{eqn:Mtx_rec2}
\overline{\br}^q_n &= \overline{\bPhi}_n(\varepsilon^q) \bA^q_n \bh^q_n + \overline{\boldsymbol{\eta}}^q_n,
\end{align}
where $\bA^q_n=[\bS^q_{n,0},...,\bS^q_{n,L_{\rm{p}}-1}]$, $\bS^q_{n,\ell}={\rm{diag}} [ (\!(\overline{\s}^q_n)\!)_{\ell} ] \in \mathbb{C}^{L_{\rm{p}}}$, 
$\bh^q_n=[(\bh^q_{n,0})^{\rm{T}},\ldots,(\bh^q_{n,L_{\rm{p}}-1})^{\rm{T}}]^{\rm{T}}$, and 
$\bh^q_{n,\ell}=[h^q[\ell, L_{\rm{cp}}+nM+\check{l}^q_{\rm{p}}],\ldots,h^q[\ell, L_{\rm{cp}}+nM+\check{l}^q_{\rm{p}}+L_{\rm{p}}-1]]^{\rm{T}}$ for $\ell=0,1,\ldots,L_{\rm{p}}-1$. Finally, $\overline{\boldsymbol{\eta}}^q_n \in \mathbb{C}^{L_{\rm{p}} \times 1}$ is the noise vector at time slot $n$. 

\begin{figure*}[ht]
 \be
 \tag{14} \label{eqn:ch2}
 \begin{aligned}
 \bOmega^q_n \!=\!\!
 \begin{bmatrix}
 h^q[0,L_{\rm{cp}}+nM+\check{l}^q_{\rm{p}}] \!&\! 
 h^q[L_{\rm{p}}\!-\!1,L_{\rm{cp}}+nM+\check{l}^q_{\rm{p}}]  \!& \!\ldots\! &\!
 h^q[1,L_{\rm{cp}}+nM+\check{l}^q_{\rm{p}}]
\\
h^q[1,L_{\rm{cp}}+nM+\check{l}^q_{\rm{p}}+1] \!&\!
h^q[0,L_{\rm{cp}}+nM+\check{l}^q_{\rm{p}}+1] \!& \!\ldots\! &\!
h^q[2,L_{\rm{cp}}+nM+\check{l}^q_{\rm{p}}+1]
\\
\vdots &\vdots &\!\!\ddots\!\! &\vdots
\\
\!h^q[L_{\rm{p}}\!\!-\!1,L_{\rm{cp}}\!+\!nM+\check{l}^q_{\rm{p}}+L_{\rm{p}}\!\!-\!1] \!&\!
h^q[L_{\rm{p}}\!\!-\!2,L_{\rm{cp}}\!+\!nM+\check{l}^q_{\rm{p}}+L_{\rm{p}}\!\!-\!1] \!& \!\ldots\! &\!
h^q[0,L_{\rm{cp}}\!\!+\!nM+\check{l}^q_{\rm{p}}+L_{\rm{p}}\!\!-\!1]
\end{bmatrix}\!\!,
\end{aligned}
\ee
\vspace{-0.4cm}
\end{figure*}

In high mobility scenarios, the channel coefficients fluctuate very rapidly. To facilitate accurate estimation in such scenarios, BEM models are used to capture the channel time variations. The CPF-BEM proposed in \cite{Muneer2015} for orthogonal frequency division multiple access (OFDMA) systems outperforms other BEM models in terms of accuracy and simplicity.
Using CPF-BEM, the channel coefficients in (\ref{eqn:ch0}) can be expressed as
\be \label{eqn:ch_bem}
h^q[\ell,\kappa]=\sum_{\gamma=0}^{\beta^q-1}B[\kappa',\gamma]c^q_{\ell}[\gamma],
\ee
where $B[.,\gamma]$ is the CPF of degree $\gamma$, $\kappa'\!=\!\frac{2 \kappa-N_{\rm{s}}+1}{N_{\rm{s}}-1}$, and $c^q_{\ell}[\gamma]$ represents the basis coefficients for user $q$.
The basis functions can be obtained in a recursive manner using $B[\kappa',\gamma+1]=2 \kappa' B[\kappa',\gamma] - B[\kappa',\gamma-1]$ with the initial conditions $B[\kappa',0]=1$ and $B[\kappa',1]=\kappa'$.
The lower bound for the number of basis functions, $\beta^q$, can be chosen based on $\beta^q \geq \lceil 2 \nu^q_{\rm{max}} T + 1 \rceil$.
Using (\ref{eqn:ch_bem}), $\bh^q_n$ can be approximated in terms of CPF coefficients, $ c^q_{\ell}[\gamma]$, as 
\be \label{eqn:bem1}
\bh^q_n=(\bI_{L_{\rm{p}}} \otimes \bB^q_n) \bc^q,
\ee
where $\bB^q_n \in \mathbb{C}^{L_{\rm{p}}^2 \times \beta^q L_{\rm{p}}}$ is the CPF-BEM matrix that stacks the values of $B[\kappa',\gamma]$ for $\kappa'\!=\!\frac{2 \kappa-N_{\rm{s}}+1}{N_{\rm{s}}-1}$, $\kappa \in \lbrace L_{\rm{cp}}\!+\!nM\!+\!\check{l}^q_{\rm{p}},\ldots, L_{\rm{cp}}\!+\!nM\!+\!\check{l}^q_{\rm{p}}\!+\!L_{\rm{p}}\!-\!1\rbrace$.
Additionally, $\bc^q=[(\bc^q_0)^{\rm{T}},\ldots,(\bc^q_{L_{\rm{p}}-1})^{\rm{T}}]^{\rm{T}} \in \mathbb{C}^{ \beta^q L_{\rm{p}} \times 1}$ for $\bc^q_{\ell}=[c^q_{\ell}(0),\ldots,c^q_{\ell}(\beta^q-1)]^{\rm{T}}$.
Substituting (\ref{eqn:bem1}) in (\ref{eqn:Mtx_rec2}), $\overline{\br}^q$ can be approximated by CPF as
\begin{align} \label{eqn:P_rec2} 
\overline{\br}^q &= \overline{\bPhi}(\varepsilon^q) \bG^q \bc^q + \overline{\boldsymbol{\eta}}^q, 
\end{align} 
where
$\bG^q\!=\![(\bG^q_0)^{\rm{T}}\!\!,\ldots,(\bG^q_{N\!-\!1})^{\rm{T}}]^{\rm{T}}$ and $\bG^q_n\!=\!\bA^q_n (\bI_{L_{\rm{p}}} \otimes \bB^q_n)$ for $n=0,1,\ldots,N-1$.

For a given set $(\bc^q,\varepsilon^q)$, the vector $\overline{\br}^q$ is assumed to have the Gaussian distribution with the mean 
$\overline{\bPhi}(\varepsilon^q) \bG^q \bc^q$ and covariance matrix $(\bm{\sigma}_\eta^q)^2 \bI_{NL_{\rm{p}}}$. Hence, the joint probability density function of $\overline{\br}^q$, parameterized by $(\tilde{\bc}^q,\tilde{\varepsilon}^q)$, is given by
\be \label{eqn:ML} f(\overline{\br}^q\!;\tilde{\bc}^q\!,\tilde{\varepsilon}^q) \!\!=\!\! \frac{1}{(\pi (\sigma^q_{\rm{\eta}})^2)^{NL_{\rm{p}}}} e^{\!\!-\frac{[\overline{\br}^q\!-\!\overline{\bPhi}(\tilde{\varepsilon}^q) \bG^q \tilde{\bc}^q]^{\! \rm{H}} \! [\overline{\br}^q\!-\!\overline{\bPhi}(\tilde{\varepsilon}^q) \bG^q \tilde{\bc}^q]}{(\sigma^q_{\rm{\eta}})^2}}\!\!.\!\! \ee
Thus, the ML estimates of the CPF coefficient vector and CFO are obtained as 
$(\hat{\bc}^q,\hat{\varepsilon}^q) = \arg \max_{\tilde{\bc}^q,\tilde{\varepsilon}^q}
\{ f(\overline{\br}^q;\tilde{\bc}^q,\tilde{\varepsilon}^q) \}$.
Taking the logarithm and removing the constant terms, the estimation problem can be simplified as 
$(\hat{\bc}^q,\hat{\varepsilon}^q) = \arg \max_{\tilde{\bc}^q,\tilde{\varepsilon}^q}
\{ g(\overline{\br}^q;\tilde{\bc}^q,\tilde{\varepsilon}^q) \}$,
where $g(\overline{\br}^q;\tilde{\bc}^q,\tilde{\varepsilon}^q)=\frac{-1}{(\sigma^q_{\rm{\eta}})^2} [\overline{\br}^q-\overline{\bPhi}(\tilde{\varepsilon}^q) \bG^q \tilde{\bc}^q]^{\rm{H}} [\overline{\br}^q-\overline{\bPhi}(\tilde{\varepsilon}^q) \bG^q \tilde{\bc}^q]$ is the joint cost function.
This maximization problem can be solved as follows.

We find $\tilde{\bm{c}}^q$ which maximizes the joint cost function parameterized by $\tilde{\varepsilon}^q$, as
\be
\tilde{\bm{c}}^q(\tilde{\varepsilon}^q) = (\bG^q)^{\dag} \bPhi^{\rm{H}}(\tilde{\varepsilon}^q) \overline{\br}^q,
\label{eqn:ch} 
\ee
where $(\bG^q)^{\dag}=\big( (\bG^q)^{\rm{H}} \bG^q \big)^{-1} (\bG^q)^{\rm{H}}$.
Then, the obtained value of $\tilde{\bm{c}}^q$ is used to find a new cost function for $\tilde{\varepsilon}^q$  which can be maximized by finding the CFO estimate through a search procedure.
Thus, we fix $\tilde{\bvar}^q$ and $\tilde{\bm{c}}^q$ that maximizes $g(\tilde{\bm{c}}^q,\tilde{\varepsilon}^q)$ can be obtained as  
\be
g_{\rm{CFO}}(\tilde{\varepsilon}^q) = (\overline{\br}^q)^{\rm{H}} \bPhi (\tilde{\varepsilon}^q) \bG^q (\bG^q)^{\dag} \bPhi^{\rm{H}}(\tilde{\varepsilon}^q) \overline{\br}^q.
\label{eqn:cost} 
\ee
Next, the CFOs for all the users are estimated by solving multiple single-dimensional search problems centered around the zero as
\be \label{eqn:cost2} 
\hat{\varepsilon}^q={\arg} \max_{\tilde{\varepsilon}^q} \{g_{\rm{CFO}}(\tilde{\varepsilon}^q)\}. 
\ee
After CFO estimation, the CPF coefficients can be determined as $\hat{\bc}^q = (\bG^q)^{\dag} \bPhi^{\rm{H}} (\hat{\varepsilon}^q) \overline{\br}^q$.
Finally, the LTV channel in the delay-time domain is estimated using (\ref{eqn:ch_bem}).

\begin{figure}[!t]
  \centering 
  {\includegraphics[scale=0.28]{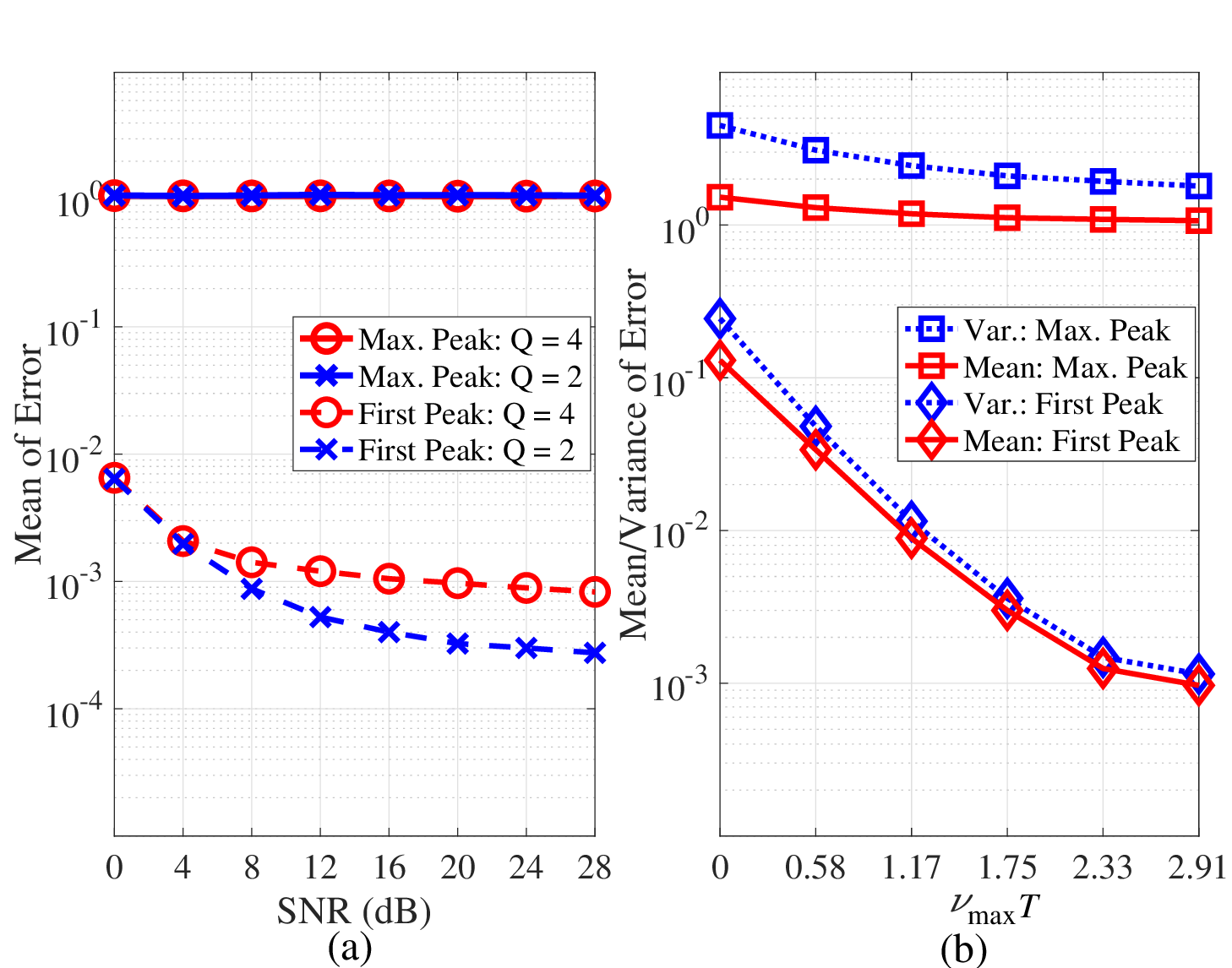}}
  \vspace{-0.3cm}
  \caption{Performance of the proposed TO estimation technique versus (a) SNR for $\nu^q_{\rm{max}} T =2.91$ and (b) normalized maximum Doppler spread at ${\rm{SNR}}=20~{\rm{dB}}$.}
  \vspace{-0.4cm}
  \label{fig:TO_SNR}
\end{figure}
\vspace{-0.1cm}
\section{Simulation Results} \label{sec:Result}
\vspace{-0.05cm}
In this section, we numerically analyze the performance of the proposed estimation techniques.
We consider an MU-OTFS uplink scenario where $M=128$ delay bins and $N=32$ Doppler bins are equally divided between $Q=2$ and $4$ users.
We use the extended vehicular A (EVA) channel model \cite{3gpp}, with length $L^q_{\rm{ch}} \leq 10$ for $q=0,\ldots,Q-1$, the bandwidth of 3.84 MHz, and the carrier frequency $f_{\rm{c}}=5.9\,\rm{GHz}$.
The power of the PCP is set to $40~{\rm{dB}}$ and our proposed pilot structure as in Fig.~2 is deployed where $l_{\rm{p}}=M-L_{\rm{p}}$ and $k^q_{\rm{p}}= q \lfloor \frac{N}{2Q} \rfloor$ for $L_{\rm{p}}=10$.
For all the users $q$, we choose the order of the CPF-BEM as $1 \leq \beta^q \leq 12$ considering the normalized maximum Doppler spreads of $0 \leq \nu^q_{\rm{max}} T \le 2.91$.
Finally, the threshold for TO estimation is chosen as $\mathcal{T}^q = 0.25, \, \forall q$. 

Fig.~\ref{fig:TO_SNR}-(a) shows the mean of TO estimation error versus SNR for $Q\!=\!2,4$.
As shown, our proposed TO estimation technique by finding the first peak of the timing metric achieves orders of magnitude higher estimation accuracy than finding the maximum peak. Our results indicate performance degradation for $Q\!=\!4$ compared to $Q\!=\!2$. This is due to the increased interference between the users' pilots as the guard between them shrinks by increasing $Q$ for fixed $M$ and $N$.
In Fig.~\ref{fig:TO_SNR}-(b), we analyze the mean and variance of the estimation error for the proposed TO estimation techniques versus normalized Doppler spread, i.e., $\nu^q_{\rm{max}} T$, for $Q\!=\!4$.
Based on our results, as Doppler spread increases, more accurate TO estimates are obtained which is due to the additional diversity gains provided by the time-selective channel.

\begin{figure}[!t]
  \centering 
  {\includegraphics[scale=0.28]{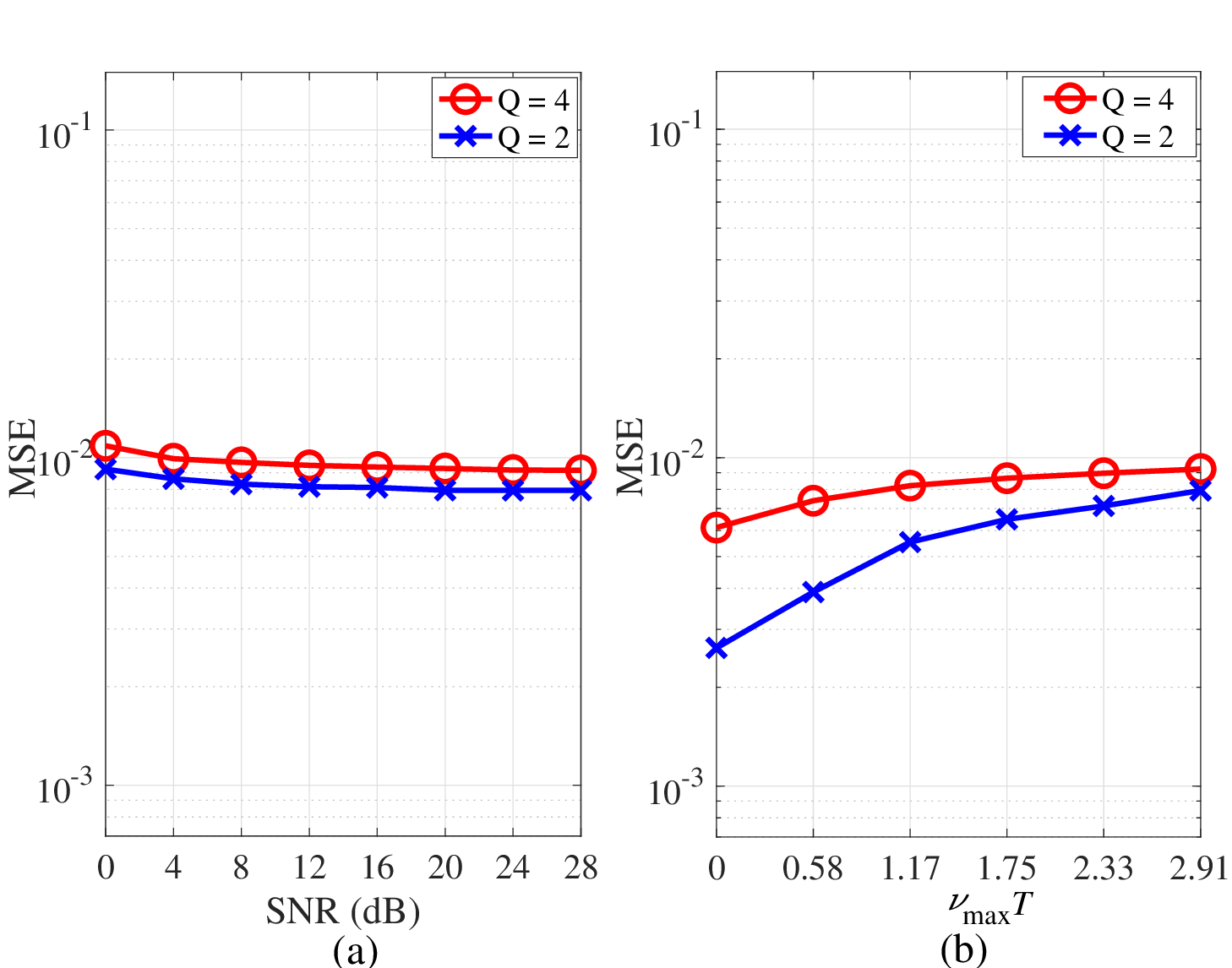}}
  \vspace{-0.3cm}
    \caption{MSE performance of the proposed CFO estimation technique versus (a) SNR for $\nu^q_{\rm{max}} T=2.91$ and (b) normalized maximum Doppler spread at ${\rm{SNR}}=20~{\rm{dB}}$.}
  \vspace{-0.4cm}
  \label{fig:CFO_SNR}
\end{figure}

To prevent interference between the users, based on the results in \cite{Raviteja2018}, a guard of at least $4 \nu^q_{\rm{max}} T$ samples is required along the Doppler dimension for the pilot region of the user $q$.
However, for $Q=4$ and $\nu_{\rm{max}}T=2.91$, this condition is not satisfied and the CFO estimation performance is degraded by multiuser interference.
In Fig.~\ref{fig:CFO_SNR}-(a), we evaluate the MSE performance of our proposed CFO estimation technique as a function of SNR. The results show that our proposed CFO estimation technique achieves an MSE in the order of $10^{-2}$, and a higher accuracy for $Q=2$ as expected.
In Fig.~\ref{fig:CFO_SNR}-(b), we examine the performance of our proposed CFO estimator with respect to normalized Doppler spread.
The results indicate that as the Doppler spread increases, the MSE performance deteriorates due to the interference between the users' pilots.

\begin{figure}[!t]
  \centering 
  {\includegraphics[scale=0.30]{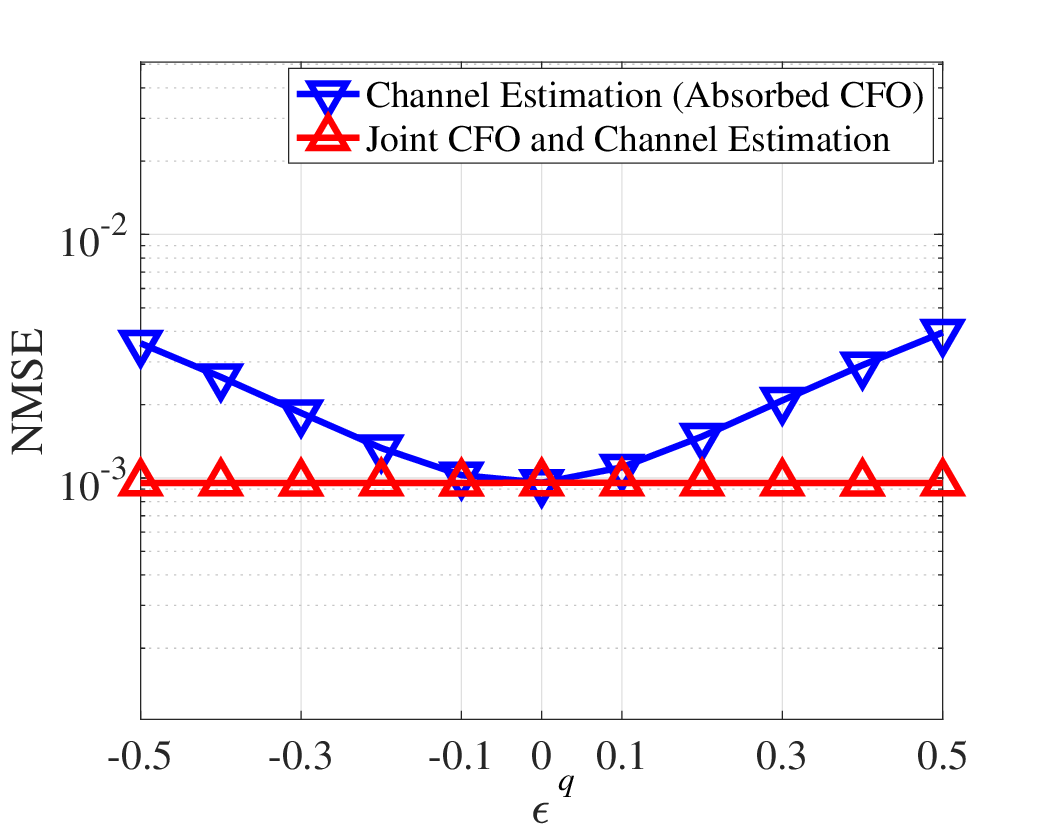}}
  \vspace{-0.3cm}
  \caption{Normalized MSE (NMSE) of the channel estimation technique.}
  \vspace{-0.4cm}
  \label{fig:Ch_CFO}
\end{figure}

Fig.~\ref{fig:Ch_CFO} illustrates the sensitivity of channel estimation to normalized CFO. When the CFO is absorbed into the channel and it is estimated as a part of the channel, the estimation accuracy of the compound channel degrades as the CFO increases. In contrast, when the CFO is separately estimated more accurate channel estimate than the joint estimation case is obtained that is not sensitive to CFO.
\vspace{-0.1cm}
\section{Conclusion}\label{sec:Conclusion}
\vspace{-0.05cm}
In this paper, for the first time, we report the problems that arise in MU-OTFS uplink if we absorb the TOs and CFOs into the channel. Specifically, the TOs need to be estimated to locate different users received pilots and then estimate the CFOs and channels. Furthermore, it is important to estimate the CFOs first and then the channels as it leads to a more accurate channel estimation performance.
We propose a pilot structure that is spectrally efficient as its CP does not need to account for the maximum TO and it deploys a shared delay region where all the users' pilots are located.
A bank of filters is used to separate the received signals of different users, and a correlation-based estimation technique detects the periodicity in the pilot signals to estimate the multiple TOs and locate the users' pilots.
Once the pilot region for each user is identified, the ML-based CFO estimation technique leverages CPF-BEM to capture variations in the LTV channel, providing CFO estimates through multiple parallel one-dimensional ML search problems.
Finally, we numerically analyzed our proposed techniques and confirmed their effective performance.

\vspace{-0.1cm}
\bibliographystyle{IEEEtran} 
\bibliography{IEEEabrv,Main}

\end{document}